\documentclass[prl,twocolumn,superscriptaddress,preprintnumbers,nofootinbib]{revtex4}
\usepackage{graphicx}
\usepackage{epsfig}
\usepackage{bm}
\usepackage{latexsym,amssymb,amsmath,amssymb,wasysym,float}

\usepackage{color}

\usepackage{scrextend}
\usepackage{mathrsfs}
\usepackage{enumitem}

\newcommand{\postscript}[2]{\setlength{\epsfxsize}{#2\hsize}
   \centerline{\epsfbox{#1}}}

\usepackage[usenames,dvipsnames]{xcolor}
\definecolor{orange}{cmyk}{0,0.5,1,0}
\definecolor{rossoCP3}{cmyk}{0,.88,.77,.40}
\definecolor{graa}{rgb}{0.8,0.8,0.8}
\definecolor{blaa}{rgb}{0.2,0.2,0.6}

\begin{document}

\title{\color{rossoCP3} Primordial Power Spectrum of Five
  Dimensional Uniform Inflation}

\author{\bf Luis A. Anchordoqui}

\affiliation{Department of Physics and Astronomy, Lehman College, City University of
  New York, NY 10468, USA
}

\affiliation{Department of Physics,
 Graduate Center,  City University of
  New York,  NY 10016, USA
}

\affiliation{Department of Astrophysics,
 American Museum of Natural History, NY
 10024, USA
}

\author{\bf Ignatios Antoniadis}

\affiliation{High Energy Physics Research Unit, Faculty of Science, Chulalongkorn University, Bangkok 1030, Thailand}

\affiliation{Laboratoire de Physique Th\'eorique et Hautes \'Energies
  - LPTHE 
Sorbonne Universit\'e, CNRS, 4 Place Jussieu, 75005 Paris, France
}

\begin{abstract}
\noindent Five dimensional (5D) uniform inflation describes a de Sitter (or
approximate) solution of 5D Einstein equations, with cosmological
constant and a 5D Planck scale $M_* \sim 10^9~{\rm GeV}$. During the
inflationary period all
dimensions (compact and non-compact) expand exponentially in terms of
the 5D proper time. This set-up requires about 40 $e$-folds to
expand the fifth dimension from the fundamental length to the micron
size. At the end of 5D inflation (or at any given moment during the
inflationary phase) one can
interpret the solution in terms of 4D fields using 4D Planck units
from the relation $M_p^2 = 2 \pi R M_*^3$, which amounts going to the
4D Einstein frame. This implies that if the compactification length
$R$ expands $N$ $e$-folds, then the 3D space would expand $3N/2$
$e$-folds as a result of a uniform 5D inflation. We reexamine the primordial power spectrum predicted by this model and show that it is
consistent with {\it Planck}'s measurements of the comic microwave
background. The best-fit to {\it Planck} data corresponds to $R \sim
10~\mu{\rm m}$. A departure of the angular power spectrum predicted by
4D cosmology is expected at multipole moment $\ell \sim 7$.  
\end{abstract}

\maketitle

Recently, we introduced the idea that a compact extra dimension
can obtain a large size by five dimensional (5D) uniform inflation, relating the
weakness of the actual gravitational force to the size of the
observable universe~\cite{Anchordoqui:2022svl}. The requirement of (approximate) flat power
spectrum of primordial density fluctuations consistent with observations of the cosmic microwave background (CMB) makes this simple idea possible only for one extra
dimension at around the micron scale~\cite{Anchordoqui:2023etp}. Thus,
this idea can be naturally combined with the dark dimension proposal for
the cosmological constant using the distance/duality conjecture within
the swampland program~\cite{Montero:2022prj}. For distances smaller
than the compactification length, 5D
uniform inflation leads to a
scale-invariant Harrison-Zel'dovich spectrum~\cite{Harrison:1969fb,Zeldovich:1972zz}, because the 2-point
function of a massless minimally coupled scalar field (such as a
slow-rolling inflaton) in de Sitter  space behaves logarithmically
at distances larger than the cosmological
horizon~\cite{Ratra:1984yq}.

For distances larger
than the compactification length (or large angles), the 5D model predicts more power
spectrum than standard 4D inflation, corresponding to a nearly
vanishing spectral index~\cite{Anchordoqui:2023etp}. Very recently,
 an investigation was carried out to ascertain the ability of future CMB experiments to constrain or detect
cosmological models that modify the CMB power spectra at large angular
scales predicted by concordance 4D
cosmology~\cite{Petretti:2024mjy}. In this Letter we reinterpret the
likelihood analysis carried out in~\cite{Petretti:2024mjy} and show
that predictions of 5D uniform inflation are consistent with CMB
measurements by the {\it Planck}
mission~\cite{Planck:2018nkj}. The best-fit to {\it Planck} data corresponds to a compactification length $\sim
10~\mu{\rm m}$. A departure of the angular power spectrum predicted by
4D cosmology is visible at multipole moment $\ell \sim 7$. 

Before proceeding, we pause to note that cosmic variance, which is large at small
multipoles (typically below an $\ell$ of order 10), limits the
precision of CMB power spectrum measurements. This is due to the
finite number of independent measurements that can be used to characterise the
temperature variance on large angular scales. These uncertainties make
it challenging to define the slope of the power spectrum of
temperature fluctuations at low
$\ell$. WMAP~\cite{WMAP:2012fli} and {\it Planck}~\cite{Planck:2018nkj} data agree with a general up-turn at low $\ell \alt
10$, but this is not necessarily a significant result, since all experiments are measuring the same universe.

It is well-known that the power spectrum of the scalar curvature
fluctuations predicted by 4D single field inflationary
models can be expressed on super-Hubble
scales as
\begin{equation}
 P_s(k) = \frac{1}{2 M_p^2 \ \varepsilon} \left(\frac{H}{2 \pi}\right)^2
 \left(\frac{k}{aH}\right)^{2\delta -\varepsilon} \,,
 \label{uno}
\end{equation}
where $k$ is the comoving momentum, $a$ is the cosmic scale factor, $M_p = 2.48 \times 10^{18}~{\rm GeV}$ is the reduced Planck
mass, $\varepsilon$ and $\delta$ are respectively the first and second
slow-roll parameters, and where $H$ characterizes the
de Sitter epoch~\cite{Riotto:2002yw}. Note that $H$ and the slow-roll
parameters are not constant, but they actually depend very slowly on time. An expansion of $\varepsilon$ and $H$ in terms of the
conformal time up to leading order
in Hubble flow parameters leads to
\begin{equation}
 P_s(k) = \frac{1}{2 M_p^2 \ \varepsilon_\circledast} \left(\frac{H_\circledast}{2 \pi}\right)^2
 \left(\frac{k}{H_\circledast}\right)^{2\delta -\varepsilon} \,,
 \label{unodos}
\end{equation}
where $H_\circledast$ and $\varepsilon_\circledast$ are the values of the Hubble parameter
and Hubble flow function at the expansion's reference point; for details
see Appendix~I. Bearing this
in mind, the spectrum of CMB anisotropies can be conveniently parametrized by
\begin{equation}
P_s(k) = A_s \left(\frac{k}{k_*} \right)^{n_s-1} \,,
\end{equation}
where $A_s \simeq 2 \times 10^{-9}$ is the scalar amplitude,
$n_s\simeq 0.96$ is the scalar spectral
tilt, and $k_* = 0.05~{\rm Mpc}^{-1}$ is the pivot scale that exits
the horizon at $N_* \equiv N_{\rm end} - 50$ e-folds from the start of
inflation ($N_{\rm start} \equiv 0$)~\cite{Planck:2018nkj}.\footnote{A point worth noting at this juncture is that any given
  present-day value ``$x$'' of a CMB scale $k$, including the pivot
  scale $k_*$, should be expressed in the form of $k= a_{\rm today} \, x~{\rm
    Mpc}^{-1}$. However, it is generally assumed that $a_{\rm today} =1$
  implicitly, and the scale is expressed as $k=x~{\rm
    Mpc}^{-1}$~\cite{Mishra:2021wkm}.} 

Now, the primordial power spectrum for 5D uniform inflation (in Planck
units) is found to be
\begin{eqnarray}
{\cal P}_s(k) & = & \frac{2 R_0 H^3}{3 \pi^3 \varepsilon} \Bigg[
\left(\frac{k}{\hat a H}\right)^{2\delta - 5 \varepsilon} S_2 (x) + \frac{\varepsilon}{3}
  \left(\frac{k}{\hat a H}\right)^{-3\varepsilon}  \nonumber \\ & \times
           &  x^2 S_4 (x) \Bigg]\, ,
\label{Pks5D}
\end{eqnarray}  
where $\hat a$ is the 5D scale factor (normalized such that $\hat a_{\rm
  start} =1$), $R_0\sim M_*^{-1}$ is the fundamental length scale (coincident with the length
of the compact dimension) at the beginning of inflation,
\begin{equation}
  S_2 (x) = {\rm coth} \ x + x \ {\rm csch}^2 x \,,
\end{equation}
and
\begin{eqnarray}
  S_4(x) & = & 15 \ {\rm coth} \ x +
  (4x^2 \ {\rm coth}^2 x + 12x \ {\rm coth}
\  x +15) \nonumber \\
  & \times &  x \ {\rm csch}^2 x  + 2x^3 \ {\rm
             csch}^4 x \, ,
\end{eqnarray}
with
\begin{equation}
x= \pi k R_0 = \frac{\pi k R}{\hat{a}_{\rm end}} = \frac{\pi k \widetilde R}{a_{\rm
    end}} = \pi k \ e^{-N} \ \widetilde{R} \,,
\label{argument}
\end{equation}
where $\hat{a}_{\rm end}$ is the 5D scale factor at the end of
inflation, $R = e^{2N/3} R_0$ is the value of the compactification length at the end
of inflation (around micron), $\tilde R = e^{N} R_0$ is the ``corrected compactification
scale'' on the brane (around kilometer), and $N$ is the number of
$e$-folds in 4D~\cite{Antoniadis:2023sya}.

Duplicating the expansion procedure on $H$ and $\varepsilon$ we
obtain
\begin{equation}
{\cal P}_s(k)  =  \frac{2 R_0 H_\circledast^3}{3 \pi^3
                    \varepsilon_\circledast} \ 
\left(\frac{k}{H_\circledast}\right)^{2\delta - 5 \varepsilon} S_2 (x),
\label{Pks5D_2}
\end{equation}  
where we have neglected the second term in (\ref{Pks5D}) because it is largely
suppressed by the slow-roll parameter when compared to the first term.

In the limit $kR_0 \gg
1$ the spectrum (\ref{Pks5D_2}) can be recast as
\begin{equation}
{\cal P}_s(k) \underset{R_0k \gg 1}{\simeq}
\frac{R_0H_\circledast^3}{3\pi^2\varepsilon_\circledast}
\biggl(\frac{k}{H_\circledast}\biggr)^{2\delta - 5 \varepsilon} \,,
\label{asym1}
\end{equation}
whereas for $kR_0 \ll 1$, the spectrum (\ref{Pks5D_2}) can be rewritten as
\begin{equation}
  {\cal P}_s (k) \underset{R_0k \ll 1}{\simeq} \frac{2H_\circledast^3}{3\pi^3\varepsilon_\circledast
     k}\ \biggl(\frac{k}{H_\circledast}\biggr)^{2\delta -
     5\varepsilon}  \, .
\label{asym2}   
\end{equation}
To a first approximation we can also ignore the pre-factor boosting the
scalar amplitude because the slow roll parameters are very small and
so the asymptotic expression (\ref{asym1}) 
gives a scale invariant spectrum
\begin{equation}
{\cal P}_{s_\gg}(k) \sim {\cal A}_{s_\gg}  \,,
\end{equation}
while (\ref{asym2}) leads to
\begin{equation}
{\cal P}_{s_\ll}(k) \sim \frac{1}{k} \ {\cal A}_{s_\ll} \,,
\end{equation}
where ${\cal A}_{s_\gg}$ and ${\cal A}_{s_\gg}$ are the corresponding
amplitudes of the two asymptotic regimes.

\begin{table}
\caption{Priors on the cosmological parameters. \label{tabla:1}}
  \begin{tabular}{c c}
    \hline
    \hline
    ~~~~~~~~~Cosmological Parameter~~~~~~~~~ & ~~~~~~~~~Priors~~~~~~~~~ \\
\hline
    $\Omega_b h^2$ & $[0.02,\, 0.0265]$\\
    $\Omega_c h^2$ & $[0.1,\, 0.135]$ \\
    $100 \theta_s$ & $[1.03,\,1.05]$\\
     $\tau_{\rm reio}$ & $[0.03,\, 0.08]$ \\
    $n_s$ & $[0.920,\,0.996]$ \\
    $\ln(10^{10} A_s)$ & $[2.763,\,4.375]$\\
   $\ln(10^{10} {\cal A}_s)$ & $[2.763,\, 4.375]$\\
    $\zeta$ & $[-6.6,\, -2.1]$ \\
            \hline
    \hline
  \end{tabular}
  \end{table}

In~\cite{Anchordoqui:2023etp} we adopted a $\Theta$-function
approximation to match the two asymptotic expressions and showed that
the resulting spectrum is partially consistent with CMB data,
including effects  of cosmic variance at large angles. However, the
use of a step $\Theta$-function approximation to accommodate the change of
behaviour amounts to throwing away any memory at smaller angles and
just parametrize the low-multipole region with an approximate $1/k$
fit.

Next, in line with our stated plan, we reinterpret the results of the likelihood analysis carried
out in~\cite{Petretti:2024mjy} using {\it Planck} 2018 CMB data on 
temperature and $E$-mode 
polarization~\cite{Planck:2018nkj}. This analysis compares predictions
from the widely accepted spatially-flat
$\Lambda$ cold dark matter (CDM) model
(supplemented by an initial stage of slow-roll inflation) and 5D uniform inflation.

$\Lambda$CDM requires only 6
independent parameters, 
\begin{equation}
\mathscr{P}_{4{\rm D}}  = \{\Omega_b h^2,\, \Omega_{c} h^2,\,
\theta_s, \, \tau_{\rm reio}, \,  n_s, \, A_s\} \,,
\label{P1}
\end{equation}
to completely specify the cosmological evolution, where $\Omega_b$
is the baryon density, $\Omega_c$ is the CDM density, $\theta_s$
is the angular size of the sound horizon at recombination, $\tau_{\rm reio}$ is the Thomson scattering
optical depth due to reionization, $n_s$ is the scalar
spectral tilt, $A_s$ is the power spectrum amplitude of adiabatic
scalar perturbations, and  $h = H_0 /(100~{\rm km} \; {\rm s^{-1}}
{\rm Mpc}^{-1})$ is the dimensionless Hubble constant. The $\Omega_i$ parameters are defined as the ratio of the present day mean density of each component $i$ to the critical density.

\begin{figure}[t]
  \postscript{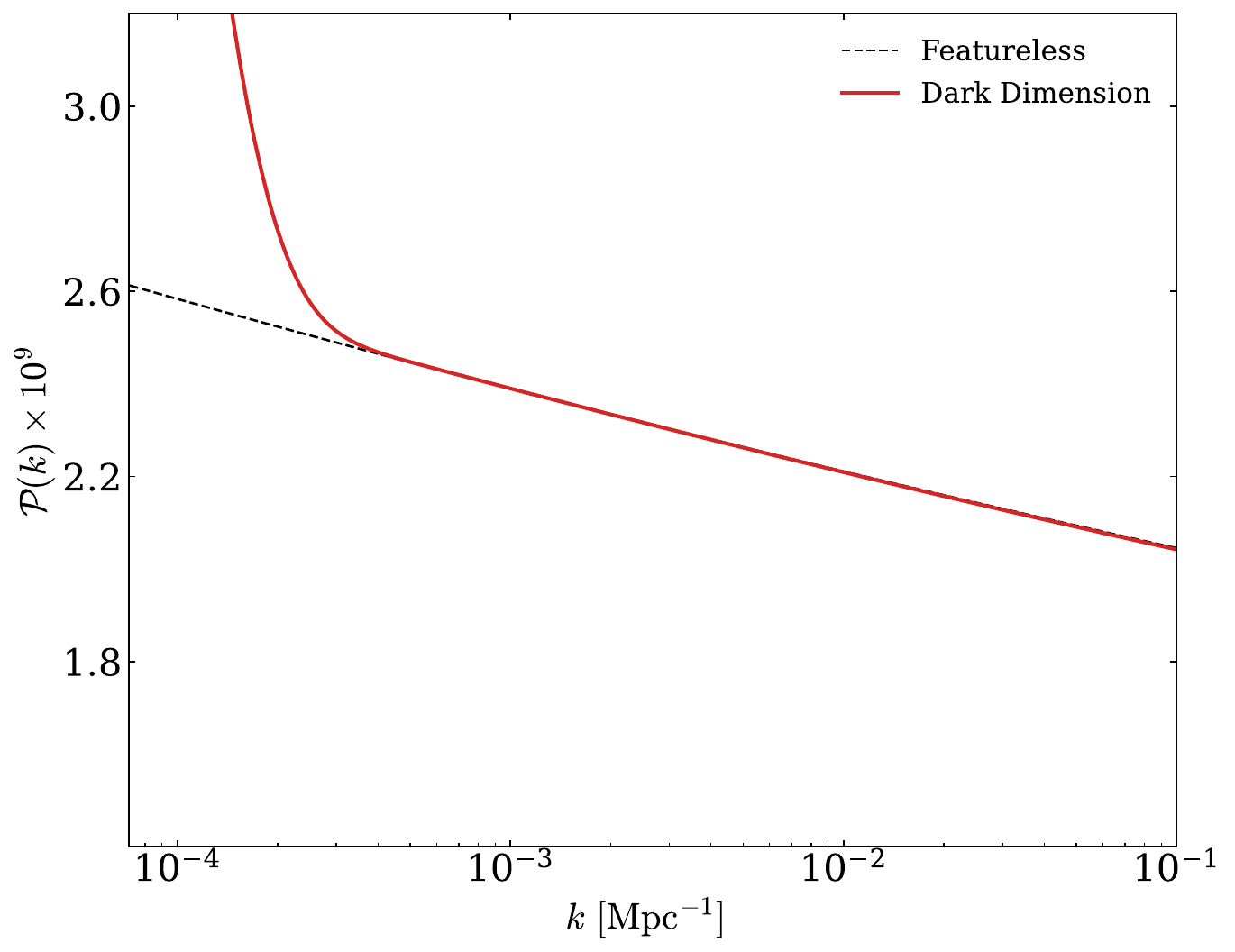}{0.9}
  \caption{Primordial power spectra of the best-fit candidates from
    the likelihood analysis of Ref.~\cite{Petretti:2024mjy}. The featureless
    best-fit (consistent with the concordance model of cosmology) is represented by the black dashed
    line and the best-fit of 5D uniform inflation by the solid line. 
    \label{fig:1}}
\end{figure}

The primordial power spectrum predicted by 5D uniform inflation given
in (\ref{Pks5D_2}) has been normalized such that $a_{\rm start} = 1$. To accomodate
the normalization of CMB data with $a_{\rm today} = 1$ we
first need to rescale the physical size of the causal patch at the beginning of
inflation $R_0$ to today ${\cal R}_{\rm today}$. After the proper
rescaling of $R_0$ has been worked out,
the primordial power spectrum of 5D uniform inflation can be parametrized by
\begin{equation}
  {\cal P}_s(k) = {\cal A}_s  \ 
  \left(\frac{k}{k_*}\right)^{n_s-1} \
    S_2( \pi k {\cal R}_{\rm today}) \, ,
\end{equation}
where 
\begin{equation}
  {\cal R}_{\rm today} = \frac{\tilde R}{a_{\rm end}} = \frac{1}{\pi k_*
    e^{\zeta} }\,,
\label{Rtoday}
\end{equation}
with
\begin{equation}
  a_{\rm end} =   \left(\frac{M_I}{10^{9}~{\rm GeV}}\right)^{-1} 2.3 \times 10^{-22}
\label{aend}
\end{equation}
the scale factor at the end of
inflation (for $a_{\rm today} =1$), $M_I$  the inflation scale, and $\zeta$ a free
parameter of the model to be determined by fitting the data. The 5D setup then requires 7 parameters,
\begin{equation}
\mathscr{P}_{5{\rm D}}  = \{\Omega_b h^2,\, \Omega_{c} h^2,\,
\theta_s, \, \tau_{\rm reio}, \,  n_s, \, A_s, \, \zeta\} \,,
\label{P2}
\end{equation}
to specify the cosmological evolution.

The analysis of~\cite{Petretti:2024mjy}  relies on uniform
priors on the cosmological parameters, which are listed in Table~\ref{tabla:1}. Note that
the priors on the $\zeta$ parameter guarantee that $S(\pi k_*
{\cal R}_{\rm today}) = 1$, and so ${\cal P}_s (k_*) = {\cal A}_s$. The
best-fit to the data gives $\zeta = -6.44$. Substituting the best-fit
value of $\zeta$
into (\ref{Rtoday})  we obtain ${\cal R}_{\rm today} \sim
4~{\rm Gpc}$. Now, substituting (\ref{aend}) into (\ref{Rtoday}) with
$M_I \sim M_* \sim 10^9~{\rm GeV}$ and ${\cal R}_{\rm today} \sim 
4~{\rm Gpc}$ leads to $\tilde R \sim
27~{\rm km}$. Converting $\tilde R$ to higher-dimensional units
it is scaled down by an additional factor $M_*/M_p$, which implies
that the compactification length at the end of
inflation is $R \sim 10~\mu{\rm m}$.

To determine whether there is an improvement in the likelihood of 5D
uniform inflation over 4D cosmology, the analysis
of~\cite{Petretti:2024mjy} reports the Bayes factor defined as the
ratio of evidences between two models. The Bayes factor is calculated
as
\begin{equation}
  \ln B = \ln {\cal Z}_{5{\rm D}} - \ln {\cal Z}_{4{\rm D}} = -1.45 \pm 0.35\,,
\end{equation}
which implies that the 5D model is statistically slightly disfavored compared
to the standard 4D scenario. Besides, 
values of $\zeta > -5.38$ are ruled out by the data at 95\%CL. This
leads to the following 95\%CL lower limits: ${\cal R}_{\rm today} > 1.4~{\rm Gpc}$
and  $R > 4~\mu{\rm m}$. 

In Fig.~\ref{fig:1} we show a comparison between the primordial power
spectra of the concordance 4D model of cosmology (dashed line) and 5D
uniform inflation (solid line) from
    the likelihood analysis of Ref.~\cite{Petretti:2024mjy}. One can
    check by inspection that the
    primordial power spectrum predicted by 5D uniform inflation departs from
    the 4D $\Lambda$CDM prediction at around $k \sim 2/{\cal R}_{\rm
      today} \sim 5 \times 10^{-4}~{\rm
   Mpc}^{-1}$. This corresponds to a multipole moment $\ell \simeq k d_A
 \simeq 7$, where $d_A = 1.4 \times 10^4~{\rm Mpc}$ is the angular
 diameter distance to the last scattering
 surface~\cite{Bridle:2003sa}. An explicit comparison between the
 different interpretations of the result from the likelihood analysis is provided in Appendix~II.

Lastly, it is constructive to connect with contrasting and
complementary perspectives to comment on two caveats of the study 
presented herein:
\begin{itemize}[noitemsep,topsep=0pt]
\item When $k$ exits the 5D horizon $R$ does not have the micron-size
  yet. We have assumed that this fact does not significantly modify the predictions of (\ref{Pks5D_2}).
\item The 5D horizon is different from
  the 4D horizon and
this may introduce an additional effect/time-dependence, which we have
assumed can be neglected.
\end{itemize}
A deeper investigation along these lines is obviously important to be done.

In summary, we have reinterpreted  the
likelihood analysis carried out in~\cite{Petretti:2024mjy} and showed 
that predictions of 5D uniform inflation are consistent with CMB
measurements by the {\it Planck} mission. In particular, the best-fit
to the data corresponds to  $R \sim
10~\mu{\rm m}$. A departure of the angular power spectrum predicted by
4D cosmology is visible at multipole moment $\ell \sim 7$. Future data
from LiteBIRD~\cite{LiteBIRD:2022cnt} and CMBS-4~\cite{CMB-S4:2016ple} will provide a decisive test for the ideas
discussed in this Letter.

\section*{Acknowledgements}

The work of L.A.A. is supported by the U.S. National Science
Foundation (NSF Grant PHY-2412679). I.A. is supported by the Second
Century Fund (C2F), Chulalongkorn University.  

\section*{Appendix~I}

An expansion of the Hubble flow functions $\varepsilon_i$ and Hubble
parameter $H$ in terms of the
conformal time $\tau = (a H)^{-1}$ around a given time
$\tau_\circledast$ leads to 
\begin{widetext}
\begin{eqnarray}
    \varepsilon_i&= &\varepsilon_i^\circledast\Bigg\{1-\ln\left(\frac{\tau}{\tau_\circledast}\right)\varepsilon_{i+1}^\circledast\big(1+\varepsilon_1^\circledast+\varepsilon_1^{\circledast2}+\varepsilon_1^\circledast\varepsilon_2^\circledast\big)
    +\frac{1}{2}\left[\ln\left(\frac{\tau}{\tau_\circledast}\right)\right]^2\varepsilon_{i+1}^\circledast\big(\varepsilon_{i+1}^\circledast+\varepsilon_{i+2}^\circledast+\varepsilon_1^\circledast\varepsilon_2^\circledast+2\varepsilon_1^\circledast\varepsilon_{i+1}^\circledast+2\varepsilon_1^\circledast\varepsilon_{i+2}^\circledast\big)
  \nonumber \\
    & & \qquad~-
        \frac{1}{6}\left[\ln\left(\frac{\tau}{\tau_\circledast}\right)\right]^3\varepsilon_{i+1}^\circledast\big(\varepsilon_{i+1}^{\circledast2}+3\varepsilon_{i+1}^\circledast\varepsilon_{i+2}^\circledast+\varepsilon_{i+2}^{\circledast2}+\varepsilon_{i+2}^\circledast\varepsilon_{i+3}^\circledast\big)\Bigg\}+\mathcal{O}\big(\varepsilon^{\circledast5}\big)
\label{app1}
\end{eqnarray}
and
\begin{eqnarray}
 H
      &= &H_\circledast\Bigg\{1+\ln\left(\frac{\tau}{\tau_\circledast}\right)\varepsilon_{1}^\circledast\big(1+\varepsilon_1^\circledast+\varepsilon_1^{\circledast2}+\varepsilon_1^\circledast\varepsilon_2^\circledast\big)
      +\frac{1}{2}\left[\ln\left(\frac{\tau}{\tau_\circledast}\right)\right]^2\varepsilon_{1}^\circledast\big(\varepsilon_1^\circledast-\varepsilon_2^\circledast+2\varepsilon_1^{\circledast2}-3\varepsilon_1^\circledast\varepsilon_2^\circledast\big)
  \nonumber \\
  &
         &\qquad~+\frac{1}{6}\left[\ln\left(\frac{\tau}{\tau_\circledast}\right)\right]^3\varepsilon_{1}^\circledast\big(\varepsilon_1^{\circledast2}-3\varepsilon_1^\circledast\varepsilon_2^\circledast+\varepsilon_{2}^{\circledast2}+\varepsilon_2^\circledast\varepsilon_3^\circledast\big)\Bigg\}+\mathcal{O}\big(\varepsilon^{\circledast4}\big)
           \, ,
\label{app2}
\end{eqnarray}
\end{widetext}  
where $\varepsilon  = \varepsilon_1$ and $\varepsilon_2 = - 2\delta +
2\varepsilon$~\cite{Antoniadis:2024abm}. Using (\ref{app1}) and (\ref{app2}) at leading
order it is easily seen that
\begin{equation}
 \frac{H^2}{\varepsilon} \left(\frac{k}{aH}\right)^{2 \delta -
   \varepsilon} \sim  \frac{H_\circledast^2}{\varepsilon_\circledast}
 \left(\frac{k}{H_\circledast}\right)^{2\delta -\varepsilon} 
\end{equation}
and that 
\begin{equation}
 \frac{H^3}{\varepsilon} \left(\frac{k}{\hat aH}\right)^{2\delta - 5\varepsilon} \sim
 \frac{H_\circledast^3}{\varepsilon_\circledast}
 \left(\frac{k}{H_\circledast}\right)^{2\delta - 5 \varepsilon} \, .
\end{equation}

\section*{Appendix II}

The discussion in this Appendix provides a critical assessment of the
likelihood analysis presented in the first version of~\cite{Petretti:2024mjy}, submitted to
the arXiv on November 2024.

In~\cite{Petretti:2024mjy} the authors first establish the relation
\begin{equation}
  R_0 = \frac{1}{\pi k_* e^{\xi}} \, ,
\label{eqApp2}
\end{equation}  
and then claim that since the length of the compact dimension at the beginning of inflation $R_0$ is
unknown, it can be determined through a
likelihood fit to the CMB data. In this Appendix we show that
this procedure leads to an ambiguity, and that actually $R_0$ cannot
be inferred from the fit.

It is well-known that in ordinary 4D inflation, where there is no extra
scale accounting for the evolution of the compact space, the
power spectrum depends on the spectral index and this gives the power
of $k/k_*$ for any reference momentum $k_*$. In other words, $k_*$ is
arbitrary. Likewise, after the end of 5D uniform inflation the compactification
length remains fixed on the micron scale and therefore a CMB
data analysis can be considered effectively 4D. However, for 5D uniform inflation, an
ambiguity emerges when backtracking the expansion of the universe
beyond the end of inflation. To understand why this is the case, we recall
that we know how $R$ grows in time from $R_0$ during inflation, but
it is important to stress that the comoving momentum $k$ is an
extra variable. Now, the relation between the times each $k$ exits the horizon
and the size of the compact space $R$ at that time is known, but
because $k$ is an extra variable we do not know at which time a given
$k$ would exit the horizon.

A crucial step in
defining (\ref{argument}) and subsequently (\ref{Rtoday}) is that the wavelength fluctuation $k$ exits the
horizon precisely at the end of inflation. This allows us to relate
the free parameter in the fit with ${\cal R}_{\rm today}$ via (\ref{Rtoday}). We reiterate that
for the best fit value $\xi = -6.44$, yielding ${\cal R}_{\rm today}
\sim 4~{\rm Gpc}$. Note that from the illogical relation (\ref{eqApp2}) it
follows that $R_0 \sim 4~{\rm Gpc}$. Clearly, the size of the compact
space $R_0 \sim M_*^{-1}$ cannot be of Gpc
scale at the beginning of inflation. We conclude that the
interpretation of the result from the fit in~\cite{Petretti:2024mjy}
using (\ref{eqApp2}) is misleading. On the other hand, assuming that
the fit in~\cite{Petretti:2024mjy}  is correct, we reinterpreted the
result using (\ref{Rtoday}), for which ${\cal R}_{\rm today}$ is of
order ${\rm Gpc}$, 
leading to a characteristic distance at the CMB epoch of ${\cal O} ({\rm
  Mpc})$ and to a compact space at the end of inflation of $\sim
10~\mu{\rm m}$. This corresponds to an angular  scale of $\sim 10^\circ$ in the
sky, for which the uncertainties in the angular power spectrum become large.


\begin{thebibliography}{99}


\bibitem{Anchordoqui:2022svl}
L.~A.~Anchordoqui, I.~Antoniadis and D.~L\"ust,
 {\color{rossoCP3} Aspects of the dark dimension in cosmology},
Phys. Rev. D \textbf{107}, no.8, 083530 (2023)
doi:10.1103/PhysRevD.107.083530
[arXiv:2212.08527 [hep-ph]].
  

\bibitem{Anchordoqui:2023etp}
L.~A.~Anchordoqui and I.~Antoniadis,
 {\color{rossoCP3}  Large extra dimensions from higher-dimensional inflation},
Phys. Rev. D \textbf{109}, no.10, 103508 (2024)
doi:10.1103/PhysRevD.109.103508
[arXiv:2310.20282 [hep-ph]].

\bibitem{Montero:2022prj}
M.~Montero, C.~Vafa and I.~Valenzuela,
{\color{rossoCP3} The dark dimension and the Swampland},
JHEP \textbf{02}, 022 (2023)
doi:10.1007/JHEP02(2023)022
[arXiv:2205.12293 [hep-th]].



\bibitem{Harrison:1969fb}
E.~R.~Harrison,
{\color{rossoCP3} Fluctuations at the threshold of classical cosmology},
Phys. Rev. D \textbf{1}, 2726-2730 (1970)
doi:10.1103/PhysRevD.1.2726

\bibitem{Zeldovich:1972zz}
Y.~B.~Zeldovich,
{\color{rossoCP3} A Hypothesis, unifying the structure and the entropy of the universe},
Mon. Not. Roy. Astron. Soc. \textbf{160}, 1P-3P (1972)
doi:10.1093/mnras/160.1.1P


\bibitem{Ratra:1984yq}
B.~Ratra,
 {\color{rossoCP3}   Restoration of spontaneously broken continuous symmetries in de Sitter space-time}
Phys. Rev. D \textbf{31}, 1931-1955 (1985)
doi:10.1103/PhysRevD.31.1931


\bibitem{Petretti:2024mjy}
C.~Petretti, M.~Braglia, X.~Chen, D.~K.~Hazra and S.~Paban,
{\color{rossoCP3}  Investigating the origin of CMB large-scale features using LiteBIRD and CMB-S4},
[arXiv:2411.03459v1 [astro-ph.CO]].

\bibitem{Planck:2018nkj}
N.~Aghanim \textit{et al.} [Planck],
 {\color{rossoCP3}  Planck 2018 results I: Overview and the cosmological legacy of Planck},
Astron. Astrophys. \textbf{641}, A1 (2020)
doi:10.1051/0004-6361/201833880
[arXiv:1807.06205 [astro-ph.CO]].

\bibitem{WMAP:2012fli}
C.~L.~Bennett \textit{et al.} [WMAP],
 {\color{rossoCP3}  Nine-year Wilkinson Microwave Anisotropy Probe (WMAP) observations: Final maps and results},
Astrophys. J. Suppl. \textbf{208}, 20 (2013)
doi:10.1088/0067-0049/208/2/20
[arXiv:1212.5225 [astro-ph.CO]].



\bibitem{Riotto:2002yw}
A.~Riotto,
 {\color{rossoCP3}  Inflation and the theory of cosmological perturbations},
ICTP Lect. Notes Ser. \textbf{14}, 317-413 (2003)
[arXiv:hep-ph/0210162 [hep-ph]].






\bibitem{Mishra:2021wkm}
S.~S.~Mishra, V.~Sahni and A.~A.~Starobinsky,
 {\color{rossoCP3}   Curing inflationary degeneracies using reheating predictions and relic gravitational waves},
JCAP \textbf{05}, 075 (2021)
doi:10.1088/1475-7516/2021/05/075
[arXiv:2101.00271 [gr-qc]].

\bibitem{Antoniadis:2023sya}
I.~Antoniadis, J.~Cunat and A.~Guillen,
 {\color{rossoCP3}    Cosmological perturbations from five-dimensional inflation},
JHEP \textbf{05}, 290 (2024)
doi:10.1007/JHEP05(2024)290
[arXiv:2311.17680 [hep-ph]].




\bibitem{Bridle:2003sa}
S.~L.~Bridle, A.~M.~Lewis, J.~Weller and G.~Efstathiou,
 {\color{rossoCP3}   Reconstructing the primordial power spectrum},
Mon. Not. Roy. Astron. Soc. \textbf{342}, L72 (2003)
doi:10.1046/j.1365-8711.2003.06807.x
[arXiv:astro-ph/0302306 [astro-ph]].


\bibitem{LiteBIRD:2022cnt}
E.~Allys \textit{et al.} [LiteBIRD],
 {\color{rossoCP3}  Probing cosmic inflation with the LiteBIRD cosmic microwave background polarization survey},''
PTEP \textbf{2023}, no.4, 042F01 (2023)
doi:10.1093/ptep/ptac150
[arXiv:2202.02773 [astro-ph.IM]].

\bibitem{CMB-S4:2016ple}
K.~N.~Abazajian \textit{et al.} [CMB-S4],
 {\color{rossoCP3}  CMB-S4 Science Book, First Edition},
[arXiv:1610.02743 [astro-ph.CO]].

\bibitem{Arkani-Hamed:1998sfv}
N.~Arkani-Hamed, S.~Dimopoulos and G.~R.~Dvali,
 {\color{rossoCP3}  Phenomenology, astrophysics and cosmology of theories with submillimeter dimensions and TeV scale quantum gravity},
Phys. Rev. D \textbf{59}, 086004 (1999)
doi:10.1103/PhysRevD.59.086004
[arXiv:hep-ph/9807344 [hep-ph]].

\bibitem{Gonzalo:2022jac}
E.~Gonzalo, M.~Montero, G.~Obied and C.~Vafa,
 {\color{rossoCP3}   Dark dimension gravitons as dark matter},
JHEP \textbf{11}, 109 (2023)
doi:10.1007/JHEP11(2023)109
[arXiv:2209.09249 [hep-ph]].

\bibitem{Macesanu:2004gf}
C.~Macesanu and M.~Trodden,
 {\color{rossoCP3}   Relaxing cosmological constraints on large extra dimensions},
Phys. Rev. D \textbf{71}, 024008 (2005)
doi:10.1103/PhysRevD.71.024008
[arXiv:hep-ph/0407231 [hep-ph]].

\bibitem{Bando:1999di}
M.~Bando, T.~Kugo, T.~Noguchi and K.~Yoshioka,
 {\color{rossoCP3}  Brane fluctuation and suppression of Kaluza-Klein mode couplings},
Phys. Rev. Lett. \textbf{83}, 3601-3604 (1999)
doi:10.1103/PhysRevLett.83.3601
[arXiv:hep-ph/9906549 [hep-ph]].


\bibitem{Bando:2000ch}
M.~Bando and T.~Noguchi,
  {\color{rossoCP3}  Brane fluctuation and new counting rules for Kaluza-Klein towers},
[arXiv:hep-ph/0011374 [hep-ph]].


\bibitem{Murayama:2001av}
H.~Murayama and J.~D.~Wells,
  {\color{rossoCP3}  Graviton emission from a soft brane},
Phys. Rev. D \textbf{65}, 056011 (2002)
doi:10.1103/PhysRevD.65.056011
[arXiv:hep-ph/0109004 [hep-ph]].

\bibitem{Kugo:1999mf}
T.~Kugo and K.~Yoshioka,
 {\color{rossoCP3}  Probing extra dimensions using Nambu-Goldstone bosons},
Nucl. Phys. B \textbf{594}, 301-328 (2001)
doi:10.1016/S0550-3213(00)00645-3
[arXiv:hep-ph/9912496 [hep-ph]].


\bibitem{Antoniadis:2024abm}
I.~Antoniadis, A.~Chatrabhuti, J.~Cunat and H.~Isono,
 {\color{rossoCP3}  New leading contributions to non-gaussianity in single field inflation},
[arXiv:2412.06616 [hep-th]].

\end{thebibliography}
\end{document}